\documentclass[12pt]{article}

\usepackage{graphicx}
\usepackage{dsfont}
\usepackage{amssymb}
\usepackage{amsmath}
\usepackage{mathrsfs}

\title{Reply to Comment on Reply to Comment on `Perfect imaging without negative refraction'}
\author{Ulf Leonhardt\\
School of Physics and Astronomy, University of St Andrews,\\
North Haugh, St Andrews KY16 9SS, UK
}
\date{\today}
\begin{document}
\maketitle

Kinsler and Favaro \cite{Comment2} point out correctly that Blaikie's numerical solution \cite{Comment} of Maxwell's equations in Maxwell's fish eye \cite{Maxwell} is causal and hence valid, a solution where no perfect image is formed. It is wrong to conclude \cite{Reply} from the existence of a causal solution with perfect imaging \cite{Fish} that one without perfect resolution is not causal; both are allowed by causality. However, Maxwell's fish eye can still perfectly image, as is shown below.

The issue whether and when Maxwell's fish eye \cite{Maxwell} images with perfect resolution can be explained by considering the propagation of light pulses, as in Kinsler's and Favaro's \cite{Comment2} numerical simulations, but with analytical techniques instead of numerics. Let us begin by writing Blaikie's numerical solution \cite{Comment} for a stationary standing wave $\widetilde{E}$ in Maxwell's fish eye in 2D (surrounded by a mirror \cite{Fish}) in terms of the analytic expressions \cite{Reply,Fish}
\begin{eqnarray}
\label{etilde}
\widetilde{E} &=& \widetilde{G}(z) - \widetilde{G}(1/z^*) \,, \quad
\widetilde{G} = \frac{P_\nu(\zeta)}{4\sin(\nu\pi)} \,,\quad \nu = \frac{1}{2}\left(\pm\sqrt{4k^2+1}-1\right) \,, \\
\zeta &=& \frac{|z'|^2-1}{|z'|^2+1} \,,\quad z' = \frac{z-z_0}{z_0^*z+1} \,.
\label{expr}
\end{eqnarray}
Here we combine the Cartesian coordinates $x$ and $y$ of the 2D fish eye in one complex number $z=x+\mathrm{i}y$; the parameter $z_0$ corresponds to the point of emission and we take as spatial unit the size $a$ of the device such that the spatial coordinates are dimensionless \cite{Fish}. The $P_\nu$ are Legendre functions (Ref.~\cite{Erdelyi}, Vol.~I) and the $\pm$ in the expression for $\nu$ refers to the sign of $k$. In the following we also measure time $t$ in units of $a/c$ with $c$ being the speed of light in vacuum. In our units the free-space wavenumber $k$ and the frequency $\omega$ are dimensionless and identical. 

We can read expression (\ref{etilde}) in two ways, as the amplitude of the stationary wave  $\widetilde{E}\exp(-\mathrm{i} k t)$ or as the Fourier transform of the light flash
\begin{equation}
\label{fourier}
E = G(z) - G(1/z^*) \,,\quad G = \int_{-\infty}^{+\infty} \widetilde{G}\, \mathrm{e}^{-\mathrm{i}kt}\, \mathrm{d}k
\end{equation}
that is emitted at point $z_0$ during one instant of time $t_0=0$. All light fields generated by a distribution of sources can be thought of as superpositions of the elementary waves $E(z,t-t_0)$ for points $z_0$ and times $t_0$ of emission. Therefore it suffices to discuss the imaging in Maxwell's fish eye considering the pulse (\ref{fourier}). The stationary wave $\widetilde{E}\exp(-\mathrm{i} k t)$ does not develop a perfect image at $z_0'=-z_0$ \cite{Reply}, but the flash (\ref{fourier}) turns out to be perfectly focused, as we show by calculating the Fourier integral (\ref{fourier}) using complex analysis.
 
The Fourier transform $\widetilde{G}(k)$ has poles on the real axis at $k_m=\sqrt{m(m+1)}$ with integer $m$ where $\sin(\nu\pi)$ is zero, branch points at $k=\pm\mathrm{i}/2$ that come from the branches of the square root in the expression (\ref{etilde}) for $\nu(k)$, and $\widetilde{G}(k)$ decays for $\mathrm{Im}\, k\rightarrow\pm\infty$. In order to establish a causal solution we move the singularities $k_m$ below the real axis by an infinitesimal amount. In this case $G$ and hence $E$ vanishes for $t<0$. For $t>0$ we extent the integration contour around the branch point $-\mathrm{i}/2$ such that $\widetilde{G}(k)$ remains on the same branch\footnote{The detour around the branch point does not contribute to the Fourier integral (\ref{fourier}) with expressions (\ref{expr}), as a consequence of the property $P_\nu$ = $P_{-\nu-1}$ of the Legendre functions \cite{Erdelyi}.}, close the integral at $\infty$ on the lower half plane and obtain from Cauchy's theorem 
\begin{equation}
\label{series}
G = \sum_{m=1}^\infty P_m(\zeta) \, (-1)^m \frac{m+1/2}{\sqrt{m(m+1)}} \, \sin\left(\sqrt{m(m+1)}\,t\right) + \frac{t}{2} \,.
\end{equation}
Close to peaks of the pulse the dominant contribution to the series (\ref{series}) comes from large-$m$ terms. There we approximate the geometric mean $\sqrt{m(m+1)}$ by the arithmetic mean $m+1/2$ and sum up the series with the help of the generating function of the Legendre polynomials $P_m$ (Eq.\ 10.10.(39) of Ref.~\cite{Erdelyi}, Vol.~II):
\begin{eqnarray}
G &\sim& \mathrm{Im} \sum_{m=1}^\infty P_m(\zeta) \,  (-1)^m \mathrm{e}^{\mathrm{i}(m+1/2)t} +\frac{t}{2} = \mathrm{Im} \frac{\displaystyle \mathrm{e}^{\mathrm{i}t/2}}{\displaystyle \sqrt{1+2\zeta\mathrm{e}^{\mathrm{i}t} + \mathrm{e}^{2\mathrm{i}t}}} -\sin\frac{t}{2} + \frac{t}{2} \nonumber\\
&\sim& \pm\,\Theta(t)\,\mathrm{Re}\frac{1}{\sqrt{-2\zeta-2\cos t}} 
\label{formula}
\end{eqnarray}
where we must take the plus sign for $t\,\mathrm{mod}\,4\pi <2\pi$ and the minus sign for $t\,\mathrm{mod}\,4\pi >2\pi$. The step function $\Theta(t)$ indicates that $G$ vanishes for $t<0$. Formula (\ref{formula}) shows that the characteristic feature of the light flash is an inverse-square-root singularity similar to the Green function of wave propagation in empty 2-dimensional space \cite{Green}.

\begin{figure}[h]
\begin{center}
\includegraphics[width=12pc]{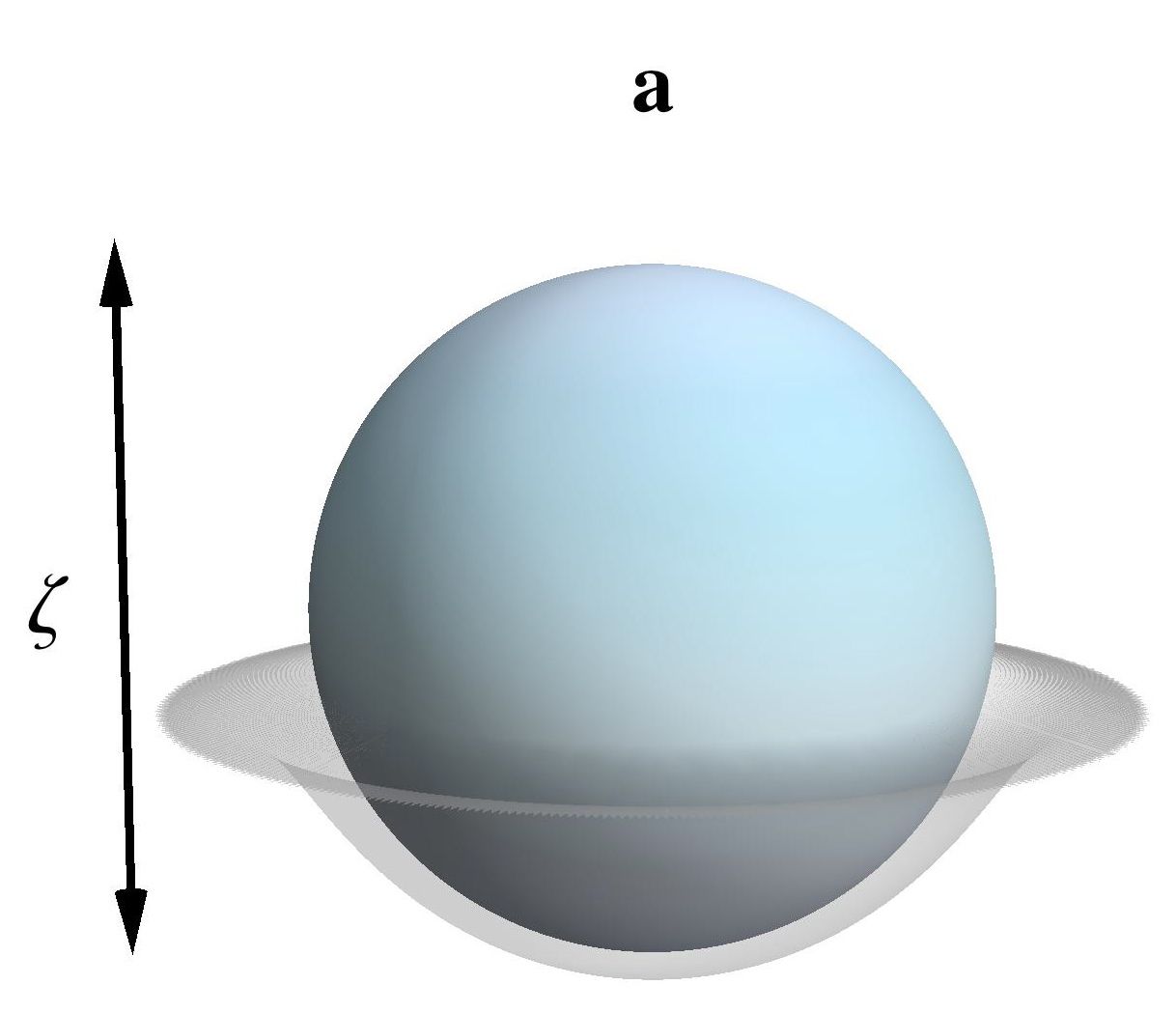}
\includegraphics[width=10pc]{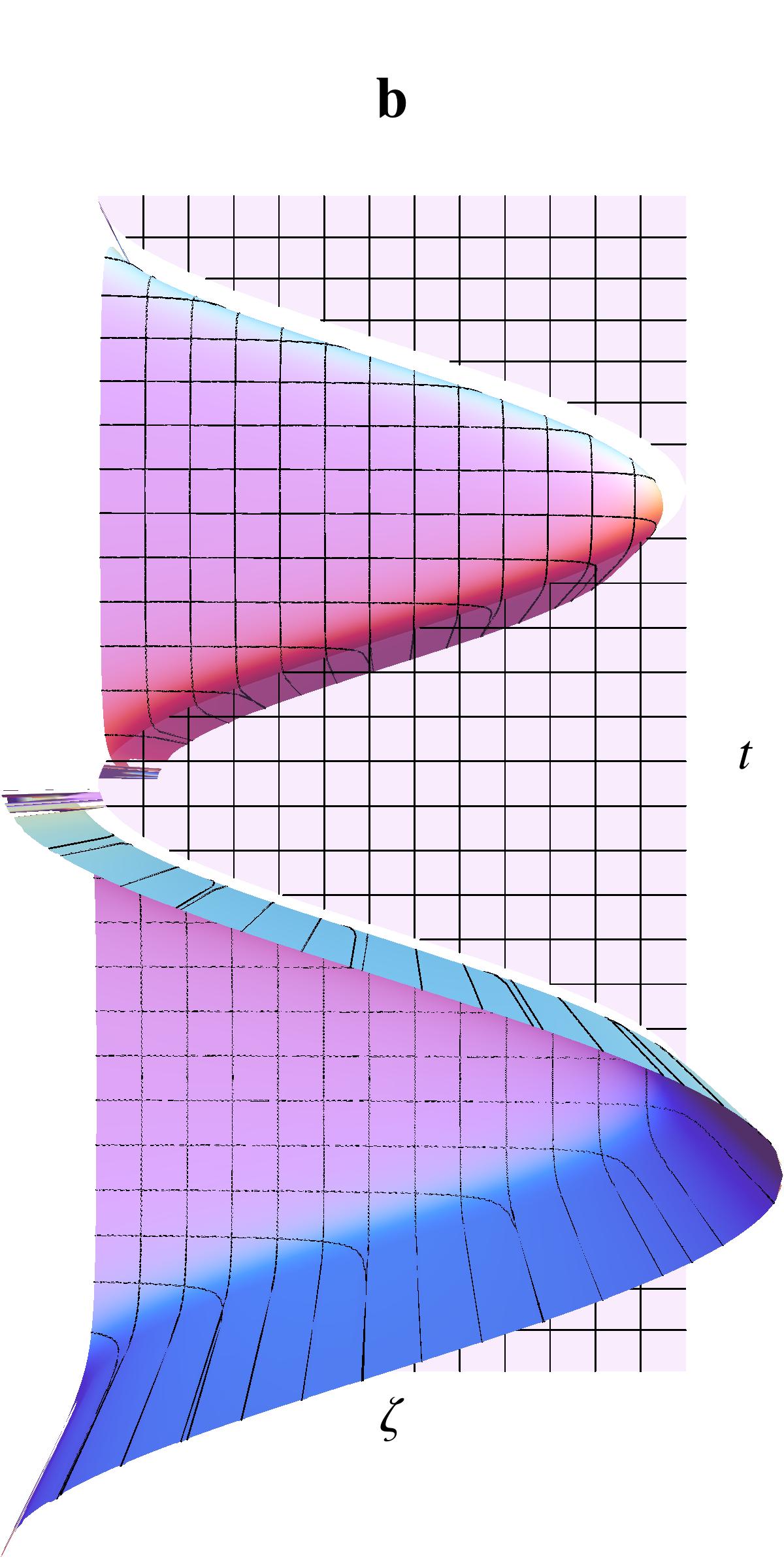}
\includegraphics[width=10pc]{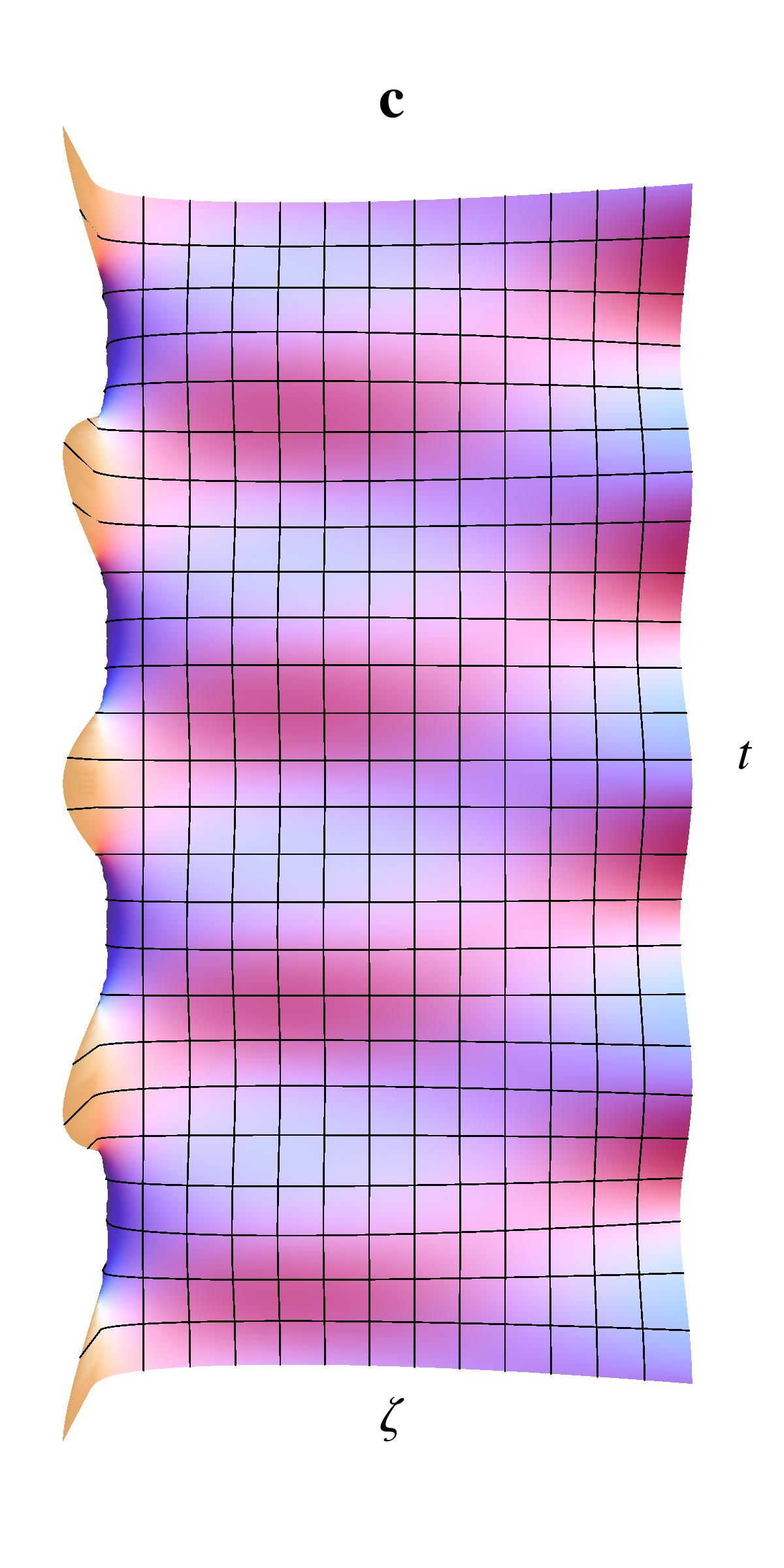}
\caption{
\small{
Light in Maxwell's fish eye \cite{Maxwell}. Light propagates in the medium of the fish eye as if it were confined to the surface of the virtual sphere shown in (a). Without loss of generality we consider a light flash emitted from the South Pole. Its wave is symmetric around the vertical axis and can only change in vertical direction $\zeta$. (b) shows a space-time diagram of the wave (\ref{formula}) illustrating the sharp feature of the flash. We see that the wave amplitude changes sign upon reflection at the source. (c) shows the stationary wave (\ref{etilde}) obtained by continuous emission (\ref{stat}) of light flashes from the source. Due to the sign change in the amplitude of the elementary flashes the image is blurred. A detector at the image point would remove the reflected wave, creating a perfect image.
}
\label{fig}}
\end{center}
\end{figure}

Figure \ref{fig} illustrates formula (\ref{formula}). The light flash is emitted at the source point $z_0$ that according to relations (\ref{expr}) corresponds to $\zeta=-1$. It propagates to the image point $-z_0$ that corresponds to $\zeta=1$ where it is reflected. The flash returns to the source point where it is reflected and changes sign. It then continues the cycle with negative sign until it changes sign again in the next reflection at the source point, and so forth. The flash thus bounces back and forth between source and image, changing sign upon reflection at the source point\footnote{In the 3-dimensional Maxwell fish eye \cite{LPfish} a light flash changes sign at the image point and not at the source point \cite{LPReply}.}. Throughout the entire propagation the dominant feature (\ref{formula}) of the light flash maintains its shape; the source that has created the flash is perfectly imaged. 

But why does the stationary wave (and Kinsler's and Favaro's long wavepackets \cite{Comment2}) not form a perfect image? It turns out that the reason is the sign change upon reflection. To see this we read the stationary wave 
\begin{equation}
\label{stat}
\widetilde{G}\,\mathrm{e}^{-\mathrm{i}kt} = \frac{1}{2\pi}\int_{-\infty}^{+\infty} G(t-t_0)\, \mathrm{e}^{-\mathrm{i}kt_0}\, \mathrm{d}t_0
\end{equation}
as the average of flashes $G(t-t_0)$ continuously emitted at times $t_0$ with phases $\exp(-\mathrm{i}kt_0)$. The wave loses its edge by averaging over alternating amplitudes, except at the source where the flashes are created from zero; the sign change causes the image to become blurred. Maxwell's fish eye has the potential of perfect imaging, but this potential is not realized yet. One essential ingredient of imaging is missing: a detector. The detector may be part of a detector array that records the image and it should only fire when it is at the correct position; ideally it should be a point detector. An ideal point detector absorbs the field at its location, acting as an outlet for the wave. The outlet eliminates the reflection back to the source and hence the sign changes that blur the image. The drain in paper \cite{Fish} is a mathematical model for a detector, which is not an artefact of the theory; on the contrary, it describes the essence of imaging. A perfect image is formed, but only when it is detected.

\section*{Acknowledgements}
Tom\'{a}\v{s} Tyc discovered by numerical summation of the series (\ref{series}) the characteristic behaviour of light flashes in the 2D Maxwell fish eye that are captured by my asymptotic formula (\ref{formula}), in particular the sign change at the source. I am grateful for discussions with him and with 
Susanne Kehr,
Yun Gui Ma,
Thomas Philbin,
and
Sahar Sahebdivan.

\end{document}